\documentclass[12pt]{article}

\usepackage{amsmath}
\usepackage{cite}
\usepackage{authblk}
\usepackage{lipsum}
\usepackage{booktabs}

\usepackage{multirow}
\usepackage{algpseudocode}
\usepackage{booktabs} 
\usepackage{multirow} 
\usepackage{tabularx} 
\usepackage{caption} 
\usepackage{graphicx} 

\title{ADLM - stega: A Universal Adaptive Token Selection Algorithm for Improving Steganographic Text Quality via Information Entropy}

\author[1]{Zezheng Qin}
\author[2]{Congcong Sun\thanks{* Corresponding author.}}
\author[3]{Taiyi He}
\author[1]{Yuke He}
\author[2]{Azizol Abdullah}
\author[2]{Normalia Samian}
\author[4]{Nuur Alifah Roslan}

\affil[1]{\footnotesize Faculty of Computer Science and Information Technology, Universiti Putra Malaysia, 43400 UPM Serdang, Selangor, Malaysia}
\affil[2]{\footnotesize Department of Communication Technology and Network, Faculty of Computer Science and Information Technology, Universiti Putra Malaysia, 43400 UPM Serdang, Selangor, Malaysia}
\affil[3]{\footnotesize Research Institute of Social Development, Southwestern University of Finance and Economics, China}
\affil[4]{\footnotesize Multimedia Department, Faculty of Computer Science and Information Technology, Universiti Putra Malaysia, 43400 UPM Serdang, Selangor, Malaysia}

\date{}

\begin{document}

\maketitle

\begin{abstract}
In the context of widespread global information sharing, information security and privacy protection have become focal points. Steganographic systems enhance information security by embedding confidential information into public carriers; however, existing generative text steganography methods face challenges in handling the long-tail distribution of candidate word pools, which impacts the imperceptibility of steganographic information. This paper proposes a quality control theory for steganographic text generation based on information entropy constraints, exploring the relationship between the imperceptibility of steganographic texts and information entropy. By controlling the information entropy of the candidate word pool within a specific range, we optimize the imperceptibility of the steganographic text. We establish upper and lower bounds for information entropy and introduce an adaptive truncation method to balance semantic coherence and lexical diversity. Experimental results demonstrate that reasonably controlling the candidate pool size and information entropy thresholds significantly enhances the quality and detection resistance of steganographic texts, showcasing broad application potential in the field of natural language processing.
\end{abstract}

\section{Introduction}
Global information sharing has become an indispensable aspect of the information society\cite{wang2024impact, azzaakiyyah2023impact}. Recently, the protection of information security and privacy has garnered increased attention. According to Shannon, there are three primary types of information security systems in cyberspace: encryption systems, privacy systems, and steganography systems\cite{yang2020behavioral}. While encryption and privacy systems safeguard the content of information, they also reveal its significance, which may introduce latent risks to the overall security framework. In contrast, steganography systems uniquely focus on embedding critical information within public carriers to obscure its presence, thereby enhancing its security\cite{abdulla2015exploiting}.

Theoretically, any carrier with redundant information space can be utilized to conceal internal secrets. This technology, known as steganography, involves embedding a secret message into public multimedia carriers—such as text, images, audio, and video—to create a steganographic carrier. This carrier can then be transmitted through public channels\cite{kessler2011overview}.

As one of the most crucial carriers of information, text plays a vital role in daily communication. Over the course of its historical development, text has evolved complex semantic coding rules aimed at improving communication efficiency, which has led to a reduction in semantic ambiguity and information redundancy. However, these factors present challenges in generating high-quality steganographic text carriers\cite{majeed2021review}.

To generate high-quality steganographic text carriers, previous studies have proposed methods for carrier selection and modification \cite{baawi2018comparative}. While these methods have addressed the complex semantic encoding rules of the original text, reducing semantic ambiguity and improving carrier quality, they fail to fully resolve the issue of low information redundancy, resulting in a lower embedding rate for secret information. To tackle this, some studies have introduced carrier generation methods that offer greater flexibility for hiding secret information and enhancing the embedding rate \cite{kumar2020recent}. However, these methods rely heavily on information-hiding algorithms to embed secret data into generated carriers, meaning the quality of the steganographic carriers largely depends on the effectiveness of information-hiding algorithms. Generally, information-hiding algorithms fall into two categories: the first employs fixed-length encoding (e.g., perfect binary tree encoding \cite{yang2018rnn}) or variable-length encoding (e.g., Huffman tree encoding \cite{yang2020vae}) with a predetermined candidate set size, where the proper set size is crucial to maintaining text quality. The second category dynamically determines the candidate set size, as seen in PPLM-stega, which addresses semantic discontinuity caused by low-probability word selection \cite{cao2022generative}. However, PPLM-stega focuses on the probability ratio of selected words, neglecting the overall probability distribution. Thus, the key challenge is determining an appropriate candidate set size while considering the overall probability distribution to generate high-quality steganographic text.

To address this challenge, we first need to understand the distribution of the candidate set. Therefore, we initially employed the Generative Large Language Model \cite{hanna2024does} to create a candidate set for a specific time step, as illustrated in Figure 1. Our observations revealed that the word distribution within the candidate set exhibited a long-tail pattern. In Figure 1(a), the distribution appears relatively flat, suggesting that a candidate set that is too restricted could limit the diversity of the generated steganographic text. Conversely, Figure 1(b) demonstrates a sharply peaked distribution, indicating a limited range of choices. In this scenario, an excessively large candidate set may result in a loss of coherence in the generated steganographic text.

\begin{figure*}[htbp]
    \centering
    \includegraphics[width=\textwidth]{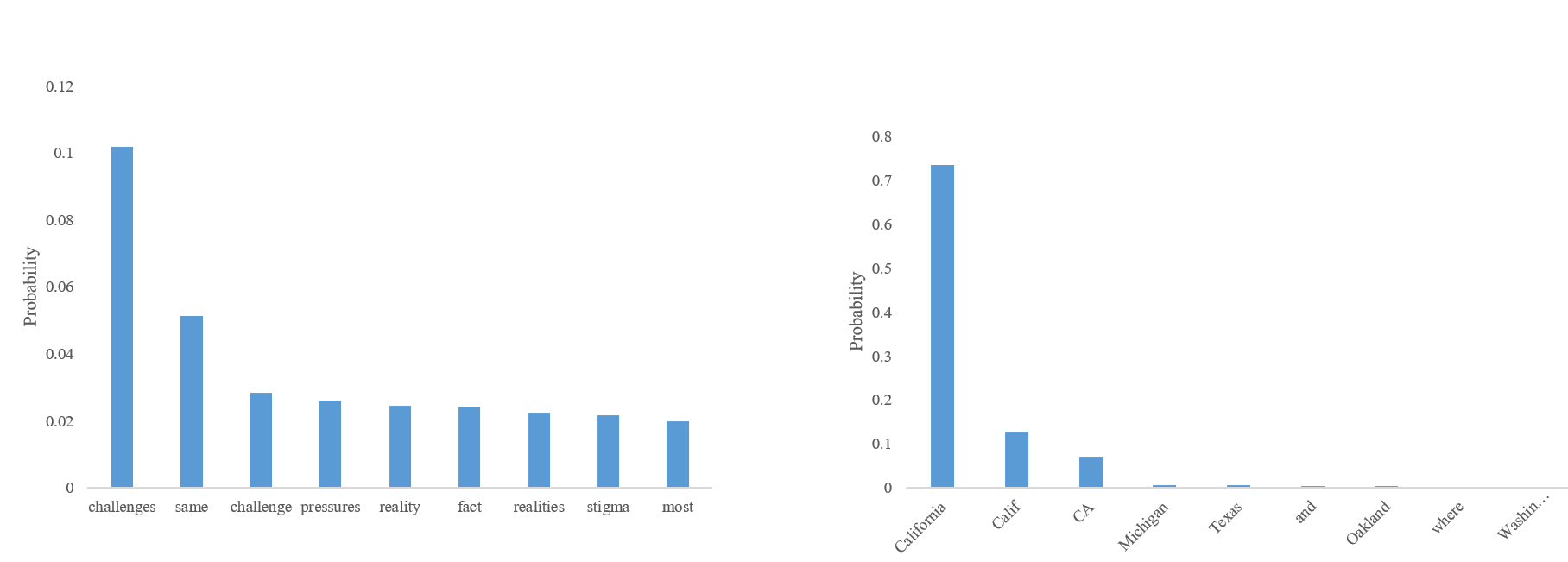} 
    \caption[Short title for list of figures]{The long-tail problem of the generative model. The underlined part is the image prediction section. Figure 1 (a) Human text prefix ``Kamala Harris was born in 1964 in Oakland, California. Her mother was a cancer research scientist from India, and her father was an economics professor from Jamaica. Growing up in a multicultural environment, Harris faced the \textunderscore as a mixed-race individual.''
    Figure 1 (b) Human text prefix ``Kamala Harris was born in 1964 in Oakland, \textunderscore.''}
    \label{fig:longtail}
\end{figure*}

Previous literature has explored various approaches to address the long-tail distribution problem. For example, the FREmax-stega\cite{pang2024fremax} adjusts the probability distribution of tokens using frequency factors to bring it closer to that of human-generated text. However, while this method increases the generation probability of low-frequency tokens within the long tail, it may not fully capture the complex distribution patterns of these tokens in human text. Certain low-frequency tokens may have higher occurrence probabilities in specific contexts, but FREmax-stega typically adjusts based on overall frequency statistics, which may not accurately reflect the distribution in these particular contexts.

Therefore, this paper introduces an adaptive word selection mechanism based on information entropy thresholds, which can address the shortcomings of handling low-frequency tokens. Unlike simple frequency-based adjustments, this approach uses information entropy to measure the system’s overall randomness and dynamically adjusts the candidate set size. This ensures that the generation probability of low-frequency tokens in specific contexts is reasonably enhanced. The mechanism not only considers the overall distribution of low-frequency tokens but also adapts to the complex patterns of specific contexts, improving the accuracy and semantic coherence of the generated steganographic text.

The specific steps for generating steganographic text are as follows. We first introduce a steganography layer before the \texttt{ADLM} output. When transmitting secret messages, this method begins by determining the prefix \( p \) of the generated carrier and setting the candidate set confidence threshold \( \epsilon \), based on a pre-agreed key between the communication parties. Concurrently, the sender encodes the secret message into a binary bitstream \( B \). During the dynamic encoding process, the steganography layer first assesses the number of candidate words after \texttt{ADLM} truncation. If the number of candidate words exceeds one, they are added to the pool of embeddable candidates. The number of bits that can be embedded is then determined by the size of this candidate pool. Finally, the word for the current time step is selected based on the encoded secret information. Experimental results demonstrate that, compared to existing text-based generation methods, the \texttt{ADLM-stega} approach produces text of superior quality, with enhanced concealment and greater resistance to detection.

The main contributions of this paper are as follows:

\begin{enumerate}
    \item This paper proposes a steganographic text generation quality control theory based on information entropy constraints, systematically exploring the quantitative relationship between the imperceptibility and semantic coherence of steganographic text. By controlling the information entropy within the candidate word pool to remain within a specific threshold range, the quality of the generated steganographic text is optimized. Additionally, the paper defines the upper and lower bounds of information entropy, $H_{\text{min}}$ and $H_{\text{max}}$, and demonstrates their association with steganographic text quality. By analyzing different probability distributions, a reasonable entropy range is proposed, providing theoretical support for generating high-quality steganographic text.  \item Based on the dynamic adjustment of information entropy, an adaptive truncation method is introduced to effectively reduce the size of the candidate word pool, ensuring a good balance between semantic coherence and lexical diversity, thereby improving the quality and effectiveness of the steganographic text. This theory not only has significant implications for steganographic text generation but can also be applied to tasks such as machine translation, dialogue generation, and automatic summarization, where controlling the information entropy of the generated text enhances its naturalness and coherence.

    \item Experimental results show that by properly controlling the size of the candidate word pool and the threshold of information entropy, the imperceptibility and semantic coherence of steganographic text can be significantly improved. Additionally, the model developed in this paper outperforms traditional methods in terms of detection resistance, and its robustness is verified through ablation experiments. This study not only provides theoretical support for steganographic text generation but also expands its potential applications in the field of natural language processing.

\end{enumerate}

\section{RELATED WORK}

\subsection{Text selection based steganography}
Li et al. (2021) proposed a method for dynamically expanding the sentiment dictionary by applying conjunction rules to identify candidate sentiment words and adding words with high similarity scores to the dictionary. Experimental results show that these conjunction rules effectively identify sentiment words, significantly reducing the likelihood of detection by statistical analysis when used for sentiment word substitution in steganography \cite{li2021text}. Mu et al. (2010) addressed challenges related to varying vocabulary, syntax, and semantics in steganographic carriers, allowing for the correction of grammatical errors and enhancing the coherence of the generated covert text by using templates \cite{munoz2010improving}. Bennett and Krista (2004) demonstrated that information can be concealed in text files by strategically placing punctuation marks, such as periods and commas, in appropriate positions \cite{bennett2004linguistic}.

Although these methods overcome the complexity of semantic encoding rules, reduce semantic ambiguity, and produce higher-quality steganographic carriers, they fail to effectively address the issue of low information redundancy in the original text, which results in a lower embedding rate for secret information.

\subsection{Text generation based steganography}
In the early stages, text generation-based steganographic techniques utilized Markov models to estimate the conditional probability distribution of each candidate word in the generated text, selecting appropriate words based on the encoded secret information \cite{yang2018automatically}. However, due to the limitations of Markov models, the generated steganographic texts only conformed to the probability characteristics of the training dataset, making them vulnerable to detection.

With advancements in Natural Language Processing (NLP), Yang et al. (2018) developed a steganographic algorithm that combines Recurrent Neural Networks (RNNs) with Huffman coding. This approach first trains a language model using an RNN and then employs Huffman coding to encode the vocabulary \cite{yang2018rnn}. Building on this foundation, Kang et al. (2020) introduced a Long Short-Term Memory (LSTM) network with keyword attention, which enhances the generation of steganographic texts by improving their semantic quality and increasing resistance to steganalysis \cite{kang2020generative}.

Furthermore, the FREmax method\cite{pang2024fremax} adjusts the probability distribution of tokens using frequency factors to make it more aligned with human-generated text. However, while this method increases the generation probability of low-frequency tokens in the long tail, it may not fully capture the complex distribution patterns of these tokens as observed in human text. In addition to these methods, other researchers have explored text generation steganography techniques based on Generative Adversarial Networks (GANs) \cite{yang2020gan} and Automatic Dialogue Systems \cite{yang2018rits}.

Although generation-based steganographic techniques can embed arbitrary secret information during text generation, these methods often fail to fully consider the overall probability distribution of the candidate pool when constructing the vocabulary, as well as the impact of the candidate pool size on the quality of the steganographic text. As a result, the generated steganographic texts may suffer from semantic ambiguity or lack coherence in vocabulary usage.

\section{The Control Theory of Imperceptibility for Steganographic Text Based on Information Entropy Bounds}

The core of steganographic text quality includes two aspects: imperceptibility and semantic coherence. Imperceptibility refers to the difficulty of detecting hidden information within the generated text, while coherence refers to the natural fluency of the text's semantics. This paper hypothesizes a quantitative relationship between the quality of steganographic text and the information entropy distribution of words in the candidate word pool. By controlling the information entropy within a specific threshold range, we can optimize text quality.

\textbf{Core Hypothesis}: The information entropy of words in the candidate word pool is closely related to the quality of steganographic text. There exists an optimal entropy value interval \([H_{\min}, H_{\max}]\) that allows the generated text to achieve the best states of imperceptibility and coherence.

In the process of generating steganographic text, information entropy \( H(X) \) is used to measure the randomness and diversity of word selection in the candidate word pool. Controlling the information entropy within a reasonable range not only aids in the semantic coherence of the generated text but also prevents excessive repetition of words or semantic dispersion during the generation process. To elaborate on how to determine \( H_{\min} \) and \( H_{\max} \), this paper will derive the process and provide detailed proofs from the perspective of information theory and probability distribution theory.

\subsection{Connection Between Information Entropy and Semantic Coherence}

Let the candidate word pool be denoted as \(\mathcal{V}\), and the probability of each word \( x \) appearing in the pool is \( p(x) \). The information entropy of the word pool is given by:
\begin{equation}
H(X) = - \sum_{x \in \mathcal{V}} p(x) \log p(x)
\end{equation}
The entropy value reflects the randomness of word selection in the candidate word pool. A higher entropy value indicates that the distribution of candidate words is more dispersed, while a lower entropy value suggests that word selection is concentrated among a few instances.

When information entropy \( H(X) \) is too high, the selection of words in the candidate word pool tends toward a uniform distribution, resulting in excessive diversity of the vocabulary distribution. In this case, the generated text may lack semantic coherence, as the generated words may be independent of one another, leading to semantic dispersion.

Conversely, when information entropy \( H(X) \) is too low, word selection becomes highly concentrated on high-frequency words, increasing the repetition rate and resulting in generated text that is stiff and lacks natural fluency.

\subsection{The Relationship Between Information Entropy and Imperceptibility}
The imperceptibility of steganographic text can be measured by the similarity between the hidden information and normal text. If the word distribution of the generated steganographic text is not significantly different from that of normal text, it becomes more difficult to detect the hidden information, resulting in better imperceptibility.

To quantify the relationship between imperceptibility and information entropy, we assume that there is some overlap between the word distribution of the candidate word pool and that of normal text $p'(x)$. During the process of generating steganographic text, by controlling the information entropy to approach the word distribution of normal text, the imperceptibility can be improved.

We assume that the word distributions of the steganographic text $p(x)$ and normal text $p'(x)$ have information entropies of $H(X)$ and $H'(X)$, respectively. To maintain good imperceptibility, the difference in information entropy between the two should be controlled within a reasonable range:
\begin{equation}
|H(X) - H'(X)| \leq \epsilon
\label{eq:entropy_difference}
\end{equation}

where $\epsilon$ is the allowable threshold for the information entropy difference. If $|H(X) - H'(X)|$ exceeds this threshold, the word distribution of the steganographic text will deviate significantly from that of normal text, making it easier to detect and thereby reducing imperceptibility.

When the information entropy $H(X)$ is too high, the word distribution tends to become uniform, meaning that the occurrence probability of each word is nearly equal. In this case, the word distribution of the steganographic text will be too scattered, inconsistent with the distribution of normal text, which increases perceptibility and reduces the effectiveness of hiding information.

Conversely, when the information entropy $H(X)$ is too low, the word distribution becomes overly concentrated, leading to the repeated appearance of high-frequency words in the generated text, reducing its fluency and naturalness. This overly concentrated word distribution is also easily detectable, thus reducing the imperceptibility of the steganographic text.

\subsection{Theoretical Derivation of Information Entropy Bounds}

To determine reasonable \( H_{\min} \) and \( H_{\max} \), we can analyze the behavior of information entropy under different distributions by examining extreme cases of probability distributions.

\paragraph{Maximum Information Entropy \( H_{\max} \) in Extreme Cases}

First, consider the case where the selection of each word in the candidate word pool is uniformly distributed, meaning that the probability of each word \( p(x) \) is equal:
\begin{equation}
p(x) = \frac{1}{|\mathcal{V}|}, \quad \forall x \in \mathcal{V}
\end{equation}
In this case, the information entropy \( H(X) \) reaches its maximum value \( H_{\max} \), which can be calculated as:
\begin{equation}
H_{\max} = - \sum_{x \in \mathcal{V}} \frac{1}{|\mathcal{V}|} \log \frac{1}{|\mathcal{V}|} = \log |\mathcal{V}|
\end{equation}
Thus, the maximum entropy value \( H_{\max} \) has a logarithmic relationship with the size of the candidate word pool \( |\mathcal{V}| \). In this scenario, the word distribution in the text is extremely dispersed, resulting in the lowest semantic coherence.

\paragraph{Minimum Information Entropy \( H_{\min} \) in Extreme Cases}

On the other hand, consider the case where the selection of words in the candidate word pool is highly concentrated, meaning that only one word has a probability close to 1, while the probabilities of other words approach 0. Let:
\begin{equation}
p(x_1) = 1, \quad p(x_i) = 0 \text{ for } i \neq 1
\end{equation}
In this case, the information entropy \( H(X) \) reaches its minimum value \( H_{\min} \):
\begin{equation}
H_{\min} = - \left(1 \log 1 + 0 \cdot \sum_{x \neq x_1} \log 0 \right) = 0
\end{equation}
In this extreme case, word selection in the text is concentrated on a single word, leading to high repetition and a lack of naturalness.

\subsection{Derivation of Reasonable Bounds for Information Entropy}

While \( H_{\max} = \log |\mathcal{V}| \) and \( H_{\min} = 0 \) represent theoretical extremes, in practical applications, excessively high or low information entropy can degrade text quality. Therefore, the reasonable entropy value interval \( [H_{\min}, H_{\max}] \) should be restricted between these extreme entropy values.

To further derive a reasonable threshold range, we assume that the semantic coherence of the generated text is inversely proportional to information entropy, while the diversity of vocabulary is positively correlated with information entropy. Specifically, we define:

\begin{itemize}
    \item \textbf{Semantic Coherence Function} \( L(H(X)) \): Represents the relationship between information entropy \( H(X) \) and the semantic coherence of the generated text, assuming \( L(H(X)) \) is a monotonically decreasing function.
    \item \textbf{Vocabulary Diversity Function} \( V(H(X)) \): Represents the relationship between information entropy \( H(X) \) and the vocabulary diversity of the generated text, assuming \( V(H(X)) \) is a monotonically increasing function.
\end{itemize}

We seek to find an optimal information entropy interval such that:
L(H(X)) \quad \text{and} \quad V(H(X))
are simultaneously balanced. To this end, we assume the existence of two critical points \( H_{\min} \) and \( H_{\max} \), which satisfy the following conditions:

\begin{itemize}
    \item When \( H(X) \leq H_{\min} \), semantic coherence is good, but vocabulary diversity is too low, resulting in high repetition in the generated text.
    \item When \( H(X) \geq H_{\max} \), vocabulary diversity is high, but semantic coherence is poor, leading to a lack of fluency in the generated text.
\end{itemize}

Thus, the optimal range of information entropy can be defined by the following equations:
\begin{equation}
H_{\min} = \alpha \log |\mathcal{V}|, \quad H_{\max} = \beta \log |\mathcal{V}|
\end{equation}
where \( \alpha \) and \( \beta \) are parameters determined based on experimental results, typically taken as \( 0 < \alpha < 1 \) and \( \alpha < \beta \leq 1 \).

\section{PROPOSED APPROACH}
In current text steganography, balancing the size of candidate pools during secret information embedding presents a significant challenge. To address this issue and enhance the quality of the generated steganographic text, improvements to the embedding algorithm are necessary. Inspired by the work of Zhu et al. \cite{zhu2024improving}, our approach involves truncating the predicted words output by the generative model based on their information entropy at each time step, using a predetermined threshold. Truncation is an algorithm that reduces the vocabulary space of a text steganography generation model to a smaller, more relevant set of words for generating the next vocabulary item. This truncation substantially impacts the quality of the generated steganographic text. Subsequently, we construct a candidate pool for embedding secret information and select the appropriate words based on the number of secret information bits and embedding rate. This method adaptively adjusts the size of the candidate pool through information entropy, effectively addressing the challenge. The methodological framework is illustrated in Figure 2.
\begin{figure*}[htbp]
    \centering
    \includegraphics[width=\textwidth]{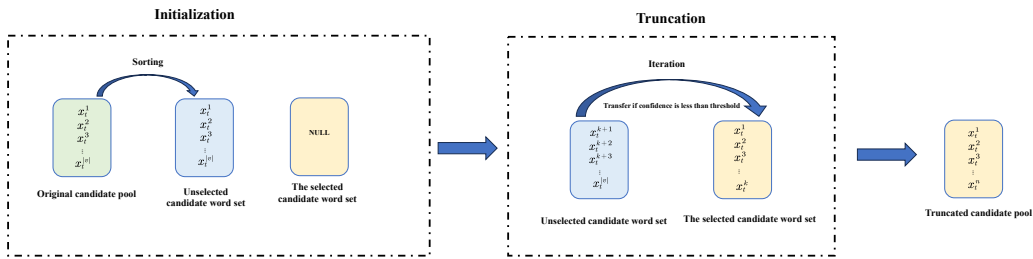} 
    \caption{Framework of the ADLM-stega approach.}
    \label{fig:example}
\end{figure*}
\subsection{The model to generate the steganographic text}
Due to the superior performance of pre-trained models in Natural Language Generation (NLG) tasks \cite{chen2019few}, this paper employs the pre-trained model GPT-2 XL \cite{keymanesh2022makes} to generate the steganographic text. 

GPT-2 XL has enhanced capabilities for text generation and understanding. Its larger parameter size enables it to produce smoother and more coherent text, and it excels in handling complex language structures and contexts, capturing subtle contextual nuances more accurately.

\subsection{Adaptive truncation of candidate set process}
This section will explain the rationale behind using information entropy to effectively truncate the size of the candidate set, describe how information entropy is employed to appropriately manage this truncation, and outline the processes of information embedding and extraction.

The candidate pool of a steganographic text generation model, when used without intervention, typically exhibits a long-tail distribution, often resulting in word repetition in the generated steganographic text \cite{pang2024fremax}. While this behavior aligns with the objective of maximizing likelihood during text generation, it does not necessarily ensure high-quality steganographic text. Additionally, the model may sample from low-probability tokens that are semantically unrelated to the already generated text, leading to incoherence. Therefore, effective truncation of the model's output words is essential for improving the quality of steganographic text generation.

To address this, we use information entropy to truncate the candidate set. Information entropy, a concept from information theory, measures the uncertainty or randomness of a random variable. Higher entropy indicates greater uncertainty, while lower entropy suggests less uncertainty \cite{kvaalseth2016measurement}. Thus, the information entropy of the candidate pool at each time step can indicate the state of information within the pool. By analyzing the information entropy across all time steps, a reasonable range of entropy typically suggests good vocabulary diversity and semantic coherence in the steganographic text. Conversely, higher or lower entropy may indicate excessive repetition or that the generated text fails to meet the criterion of imperceptibility \cite{kvaalseth2016measurement}.

For the discrete variable distribution of the candidate pool at each time step.
For each time step, the information entropy$\mathrm{H}(X)$, of the model output words discrete variable distribution is\begin{equation}H(X)\leq\log|\mathcal{V}|\end{equation}

In order to reduce the impact of the vocabulary dimension of the text generation model, we use the minimum-maximum method to normalize the entropy. For each time step of the model output words discrete variable distribution $X$, the confidence $Conf(X)$ is defined as the minimum-maximum scale of entropy, and its value range is [0,1]:\begin{equation}\mathrm{Conf}(X)=1+\frac{\sum_{x\in\mathcal{V}}p(x)\log p(x)}{\log|\mathcal{V}|}.\end{equation}

$\sum_{x\in\mathcal{V}}p(x)\log p(x)$ can be divided into two parts. The first part is $\sum_i^kp_i\operatorname{log}p_i$, where $k$ represents the number of model output words selected in the candidate pool at the current time step, which is 0 at the beginning. The second part is the unknown part, which introduces the maximum uncertainty and is expressed as $(1-\sum_i^kp_i)\log\frac{1-\sum_i^kp_i}{|\mathcal{V}|-k}$. That is, the maximum influence of the unselected words in the candidate pool at the current time step.

In order to determine the reasonable size of the candidate pool, we define the candidate pool size at the current time step as $k$ and the confidence state as\begin{equation}\begin{aligned}\operatorname{Conf}_k(X)&=1+\frac1{\log|\mathcal{V}|}(\sum_i^kp_i\log p_i\\&+(1-\sum_i^kp_i)\log\frac{1-\sum_i^kp_i}{|\mathcal{V}|-k}),\end{aligned}\end{equation}
where $P_i$ is the abbreviation of $p(x_i)$.

Based on the previously defined optimal information entropy interval
the confidence can be mapped to the range of information entropy.

Firstly, the optimal interval for information entropy is located between
$
H_{\text{min}} \quad \text{and} \quad H_{\text{max}} \quad \text{as follows:} 
$
\begin{equation}
\alpha \log |\mathcal{V}| \leq H(X) \leq \beta \log |\mathcal{V}|
\end{equation}

This interval can be mapped to the range of confidence. Since the confidence \( \mathrm{Conf}(X) \) is based on the normalized value of information entropy, we can define the corresponding confidence interval. We know that the information entropy satisfies \( H(X) \leq \log |\mathcal{V}| \) and that the range of \( \mathrm{Conf}(X) \) is \([0, 1]\).

Therefore, when the information entropy is within the bounds \( H_{\text{min}} \) and \( H_{\text{max}} \), the corresponding confidence interval can be represented as:
\begin{equation}
\mathrm{Conf}_{\text{min}} = 1 + \frac{\alpha - 1}{1}, \quad \mathrm{Conf}_{\text{max}} = 1 + \frac{\beta - 1}{1}
\end{equation}

Thus, we can obtain:
\begin{equation}
\alpha = \mathrm{Conf}_{\text{min}}, \quad \beta = \mathrm{Conf}_{\text{max}}
\end{equation}

Next, we can update the previous formulas, expressing \( \alpha \) and \( \beta \) in terms of confidence. The final expressions are:
\begin{equation}
H_{\text{min}} = \mathrm{Conf}_{\text{min}} \log |\mathcal{V}|
\end{equation}
\begin{equation}
H_{\text{max}} = \mathrm{Conf}_{\text{max}} \log |\mathcal{V}|
\end{equation}

We can adjust the parameter Conf to determine the relationship between confidence and information entropy, thereby controlling the semantic coherence and vocabulary diversity of the generated text.

After that, We define whether a candidate word is selected by the candidate pool at each time step as the degree of influence of selecting or not selecting the candidate word on the generation of the candidate pool. Specifically, it is defined as the difference between adjacent candidate words
\begin{equation}
\begin{split}
\Delta\mathrm{Conf} &= \frac{1}{\operatorname{log}|\mathcal{V}|}(\mathrm{Conf}_{k}(X)-\mathrm{Conf}_{k-1}(X)) \\
&= \frac{1}{\log|\mathcal{V}|}(p_k\log p_k+(1-\sum_i^kp_i)\log\frac{(1-\sum_i^kp_i)}{|\mathcal{V}|-k} \\
&\quad -(1-\sum_i^{k-1}p_i)\log\frac{(1-\sum_i^{k-1}p_i)}{|\mathcal{V}|-k+1}).
\end{split}
\end{equation}

We find that $\Delta\mathrm{Conf}$ is always greater than or equal to 0. This shows that increasing the number of candidate words in the candidate pool at each time step will improve the quality of the stego text to a certain extent.

Also\begin{equation}
\Delta \text{Conf} \leq \left(1 - \frac{P_{k-1}}{i \cdot p_i}\right) \log |V| \cdot \log(|V| - k + 1 )
\end{equation}

As the known candidate words increase, a decrease in $p_k$ can be observed. In addition, $log(|V| - k + 1)$ and $(1-\sum_i^{k-1}p_i)$ both show a decrease. Eventually the upper and lower bounds approach $0$. This shows that when the number of candidate words exceeds a certain limit, the quality of the generated steganographic carrier will be reduced.

According to the above principles, the steps for using information entropy to determine the size of the candidate pool for each time step are as follows. As shown in Figure 3.

First, initialize the vocabulary, set the set of candidate words selected by the candidate pool at the current time step to be $\mathcal{A}_{\boldsymbol{x}<t}$, and define the set of unselected candidate words as $\mathcal{B}_{\boldsymbol{x}<t}$. Initialization includes setting $\mathcal{A}_{\boldsymbol{x}<t}$ to an empty set and setting $\mathcal{B}_{\boldsymbol{x}<t}$ to an ordered set of the entire vocabulary. The order is based on the descending probability predicted by the text generation model.

The model vocabulary is then truncated, which can be viewed as an iterative process involving transferring candidate words from $\mathcal{B}_{\boldsymbol{x}<t}$ to $\mathcal{A}_{\boldsymbol{x}<t}$ with the highest probability. Subsequently, we calculate the incremental change in the $Conf$ metric. The iteration continues until the plausibility is below a preset threshold, denoted as $\boldsymbol{\epsilon}$. At this point, we determine $\mathcal{A}_{\boldsymbol{x}<t}$ as the truncation space.

\begin{figure*}[htbp]
    \centering
    \includegraphics[width=0.75\textwidth]{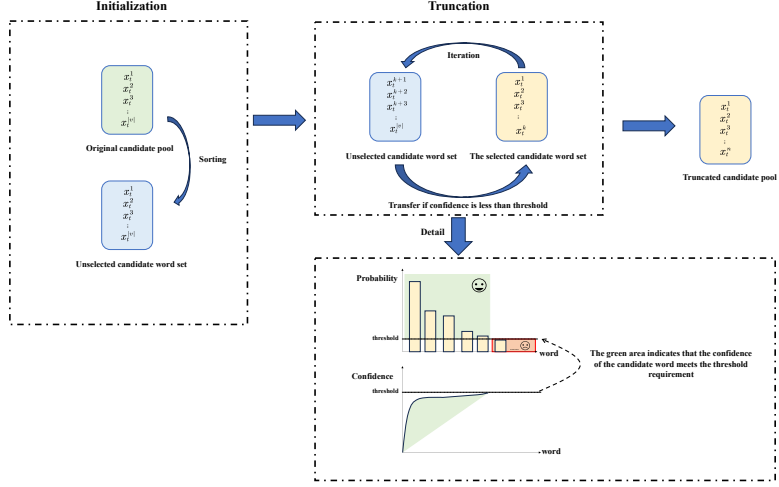}
    \caption{Candidate set truncation process.}
    \label{fig:example}
\end{figure*}

\subsection{Secret Information Embedding Algorithm}

The ADLM-stega method proposed in this paper embeds secret messages during the text generation process. The core idea is to encode candidate words at each step and determine the length of the binary bit stream that can be embedded based on the size of the candidate pool. The corresponding code word is then selected according to this binary bit stream. The specific process is as follows:

\begin{enumerate} 
    \item The sender determines the prefix \( p \) and the threshold \( \varepsilon \) based on the pre-agreed key. The sender encodes the secret message into a binary bit stream \( B \).
    
    \item The generative text model performs the text generation task using the prefix \( p \).
    
    \item The sender establishes a candidate pool. If \( n = 1 \), it cannot embed the secret message, so the sender updates \( p = p + x_t \) and returns to Step 2. If \( n > 1 \), the sender adds all truncated candidate words to the pool.
    
    \item The sender determines the length of the binary bit stream that can be embedded based on the candidate pool size. Let \( m \) denote the pool size at time \( t \) (where \( m > 1 \)). The bit flow rate is \( k = \lfloor \log_2 m \rfloor \). The sender compares the remaining unembedded message length \( l_{rh} \) with \( k \). If \( l_{rh} \geq k \), the first \( 2^k \) candidate words are selected based on probability and encoded using a binary tree. If \( l_{rh} < k \), the first \( 2^{l_{rh}} \) candidate words are selected and encoded. Finally, the word \( x_t \) is chosen according to the binary bit stream.
    
    \item Repeat Steps 2-4 until all binary bit streams are embedded.
\end{enumerate}

\begin{figure}
\centering
\includegraphics[width=\textwidth]{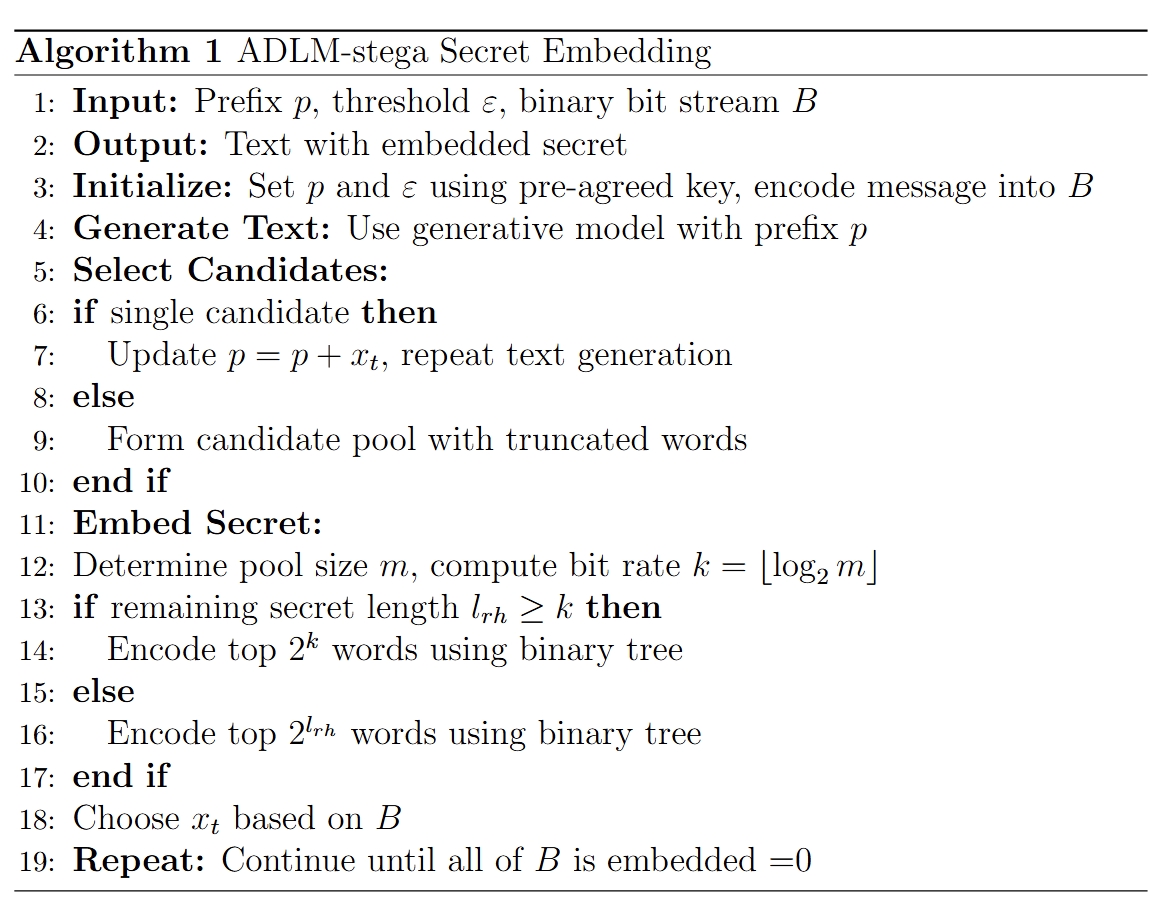}
\caption*{}
\end{figure}

\subsection{Secret Information Extraction Algorithm}

During the hidden text extraction process, the receiver uses the same model and key to extract the encoded secret message. The specific process is as follows:

\begin{enumerate} \item The receiver determines the prefix 
\( p \) and threshold \( \epsilon \) from the shared key and treats the hidden text as a word sequence.

\item The generative text model generates text with prefix \( p \) and threshold \( \epsilon \).

\item The receiver creates a candidate word pool as the sender does.

\item The receiver determines the binary bit stream segment length \( k \) based on the pool size, selects and encodes words, and updates the prefix \( p \) accordingly.

\item The process is repeated until all secret messages are extracted.
\end{enumerate}

\section{EXPERIMENTAL RESULTS AND ANALYSIS}
In this section, we begin by conducting experiments and analyses to evaluate the quality of the generated steganographic text. Subsequently, we investigate the relationship between the size of the candidate pool and the threshold. We then assess the resistance of our method to steganalysis. Finally, we perform ablation studies on the model to further validate its robustness.

The experiments are conducted using the GPT-2XL model for generating steganographic texts. The proposed method is implemented in PyTorch and executed on an RTX 3080 GPU.

\subsection{The imperceptibility and semantic coherence of steganographic text.}

First, the regulation of information entropy \( H(X) \) during the steganographic text generation process significantly impacts text quality. Theoretically, after controlling with information entropy thresholds, the generated text should approach optimal values in perplexity \( PPL \) and vocabulary distribution \( D \), specifically:

\begin{equation}
PPL \approx PPL_{\text{min}}, \quad D \approx D_{\text{opt}}
\end{equation}

The objective of the experiments is to compare the perplexity and vocabulary distribution of the text generated within different ranges of information entropy \( H(X) \) by adjusting the size of the candidate word pool \( V \), thereby validating the effectiveness of this theory.

To validate the effectiveness of the theory, a series of experiments were designed to calculate information entropy using different sizes of the candidate word pool and generate text. We evaluated the performance of the generated text based on two core indicators:

\begin{itemize}
    \item Perplexity \( PPL \): This reflects the semantic coherence of the text. Ideally, the perplexity of the generated text should be close to the optimal value \( PPL_{\text{min}} \).
    \item Vocabulary distribution \( D \): This measures the vocabulary diversity of the text. The optimal vocabulary distribution \( D_{\text{opt}} \) represents high-quality text generation.
\end{itemize}

The specific steps are as follows:

\begin{enumerate}
    \item Select candidate word pools \( V \) of different sizes, calculate the variation in information entropy \( H(X) \), and generate steganographic text.
    \item Compare the quality of the generated text when \( H(X) \) falls within a preset reasonable range \( [H_{\text{min}}, H_{\text{max}}] \) with the differences observed when not within that range, focusing on the perplexity and vocabulary distribution of the generated text.
\end{enumerate}

In the experiments, different steganographic text generation models were used, including our proposed model based on information entropy regulation (ADLM-stega), as well as traditional RNN-stega and PPLM-stega models, generating 1,000 steganographic text samples each. We assessed the quality of the text based on perplexity (PPL) and Distinct metrics.

As shown in Table 1, our model maintains a relatively low perplexity level despite an increase in embedding rate, ensuring the semantic coherence of the text. At the same time, the lower values of the Distinct metric indicate that the randomness and vocabulary diversity during the text generation process are well controlled.

\begin{table}[htbp]
  \centering
  \caption{Performance comparison of different models}
  \label{tab:addlabel}
  \footnotesize 
  \renewcommand{\arraystretch}{1.2} 
  \begin{tabularx}{\linewidth}{>{\centering\arraybackslash}X >{\centering\arraybackslash}X cccc} 
    \toprule
    \multirow{2}{*}{Model} & \multirow{2}{*}{Index} & \multicolumn{4}{c}{BPW} \\
    \cmidrule{3-6}          &       & 1     & 2     & 3     & 4 \\
    \midrule
    \multirow{2}{*}{Ours} & PPL     & 36.73  & 44.50 & 62.27  & 85.21  \\
                          & Distinct & 0.207 & 0.210 & 0.211  & 0.213  \\
    \multirow{2}{*}{RNN-stega} & PPL     & 1593.47 & 445.61 & 841.62 & 972.92 \\
                               & Distinct & 0.352 & 0.305 & 0.263 & 0.303 \\
    \multirow{2}{*}{PPLM-stega} & PPL     & 94.17  & 156.54 & 245.78 & 368.64 \\
                               & Distinct & 0.339 & 0.347 & 0.327 & 0.334 \\
    \bottomrule
  \end{tabularx}%
\end{table}

\begin{figure}[h]
    \centering
    \includegraphics[width=0.75\linewidth]{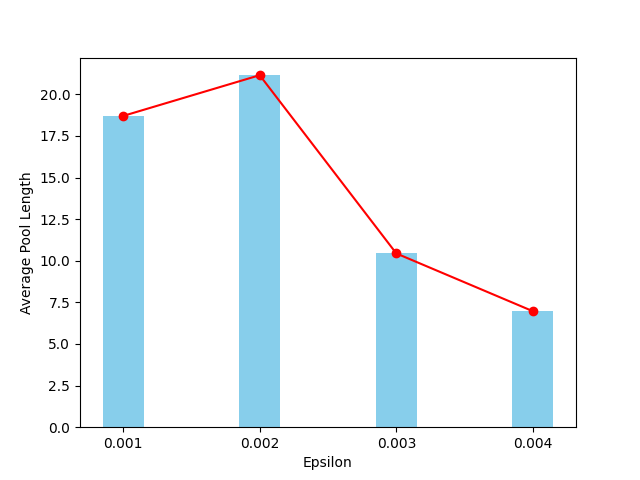} 
    \caption{Relationship between the size of the candidate pool and the threshold}
    \label{fig:example}
\end{figure}

\begin{table}[htbp]
  \centering
  \scriptsize 
  \caption{Comparison of anti steganography performance}
    \begin{tabular}{ccccccc}
    \toprule
    \multirow{2}[4]{*}{Bpw} & \multicolumn{2}{c}{RNN-stega} & \multicolumn{2}{c}{PPLM-stega} & \multicolumn{2}{c}{Ours} \\
    \cmidrule{2-7}          & ACC   & R & ACC   & R & ACC   & R \\
    \midrule
    1     & 0.7125 & 0.445 & 0.775 & 0.575 & 0.6825 & 0.505 \\
    2     & 0.765 & 0.54  & 0.685 & 0.37  & 0.7075 & 0.585 \\
    3     & 0.7375 & 0.51  & 0.7825 & 0.61  & 0.685 & 0.605 \\
    4     & 0.7225 & 0.51  & 0.7675 & 0.59  & 0.655 & 0.555 \\
    5     & 0.725 & 0.51  & 0.765 & 0.595 & 0.685 & 0.595 \\
    6     & 0.7   & 0.51  & 0.76  & 0.58  & 0.62  & 0.535 \\
    7     & 0.7075 & 0.515 & 0.7525 & 0.56  & 0.6425 & 0.57 \\
    8     & 0.7025 & 0.535 & 0.75  & 0.615 & 0.625 & 0.56 \\
    \bottomrule
    \end{tabular}%
  \label{tab:addlabel}%
\end{table}%

\begin{table}[htbp]
  \centering
  \caption{Comparison of Model Performance in Ablation Experiments}
  \label{tab:addlabel}
  \scriptsize 
  \begin{tabular}{p{2cm}p{0.75cm}p{0.75cm}p{0.75cm}p{0.75cm}p{0.75cm}}
    \toprule
    \multirow{2}[4]{*}{Model} & \multirow{2}[4]{*}{Index} & \multicolumn{4}{c}{BPW} \\
    \cmidrule{3-6}          &       & 1     & 2     & 3     & 4 \\
    \midrule
    \multirow{2}[1]{*}{Original model} & PPL   & 36.73 & 44.5  & 62.27 & 85.21 \\
    & Distinct & 0.207 & 0.21  & 0.211 & 0.213 \\
    \multirow{2}[1]{*}{Ablation model} & PPL   & 1005.24 & 1153.37 & 1257.62 & 1671.34 \\
    & Distinct & 0.329 & 0.31  & 0.305 & 0.294 \\
    \bottomrule
  \end{tabular}%
\end{table}

\subsection{Relationship Between the Size of the Candidate Pool and the Threshold}

The size of the candidate pool in the model is influenced by the threshold value. To explore the relationship between candidate pool size and threshold, we generated 200 samples with varying thresholds and calculated the average candidate pool size for each threshold. The results are presented in Figure 4. 

We observe that the average candidate pool size remains relatively stable when the threshold is set to 0.001 and 0.002, but decreases rapidly as the threshold increases beyond these values. This behavior indicates that the distribution of the candidate pool follows a long-tail distribution pattern.

\subsection{Anti-Detectability}

The primary objective of steganography is to securely transmit secret information, making it essential to evaluate the anti-detection capability of the proposed method \cite{wu2021linguistic}. We assess anti-detection ability using accuracy (ACC) and recall (R) metrics for the steganalysis detection method. 

From the results presented in Table 2, we draw the following conclusions. First, as the embedding rate increases, the steganographic text becomes more detectable. Second, our method demonstrates superior anti-detection ability compared to the other two methods. Third, the steganalysis results further corroborate the findings from the imperceptibility experiments.

\subsection{Ablation Experiment}
\begin{figure}[h]
    \centering
    \includegraphics[width=\linewidth]{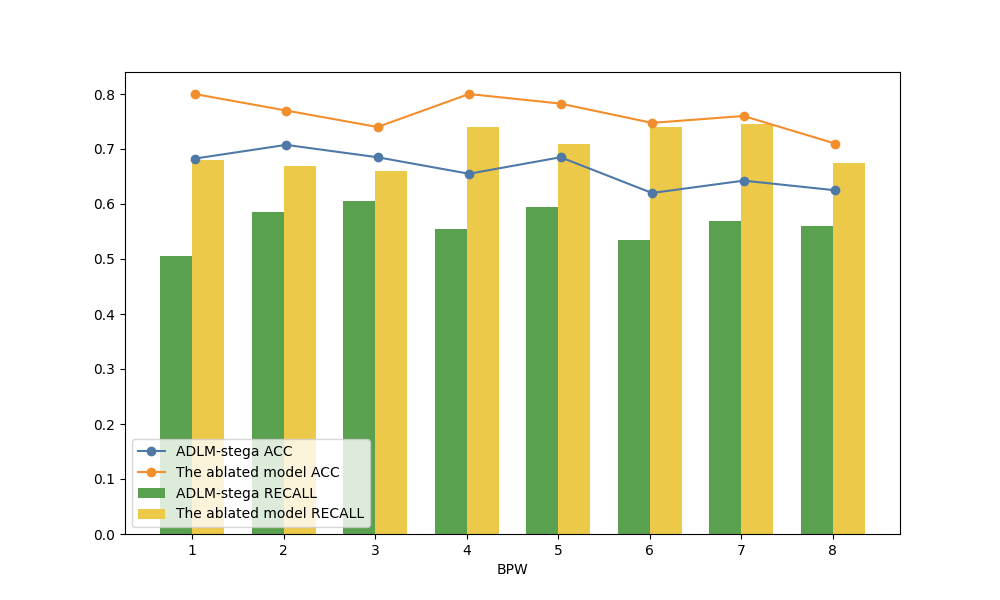} 
    \caption{Results of the Ablation Experiment.}
    \label{fig:example}
\end{figure}
To assess the robustness of our model, we performed ablation experiments by removing the confidence module and evaluating the quality of the steganographic text and the anti-steganography capabilities. We compared these results with those of the original model. The experimental results are presented in Figure 5 and Table 3.

The findings reveal the following: first, the imperceptibility of the model without the confidence module is generally lower than that of the original model. Second, the anti-steganography capability of the ablated model is significantly reduced compared to the original model. These results confirm that the inclusion of the confidence module is crucial for maintaining the robustness of our model.

\section{Conclusion}

The ADLM-stega method proposed in this paper effectively improves the quality of stego-texts by adaptively adjusting the candidate set size, addressing the shortcomings of existing methods in handling long-tail distribution issues. Firstly, we demonstrate the impact of candidate set size on the quality of stego-texts, proving that an appropriate candidate set size is crucial for generating high-quality stego-texts. Secondly, we analyze the relationship between the candidate pool size and the truncation algorithm threshold, revealing the long-tail effect in text steganography. Finally, ADLM-stega excels in terms of concealment and resistance to detection, validating its effectiveness in stego-text generation. Experimental results show that this method can generate semantically consistent and naturally flowing stego-texts across various embedding rates, with strong detection resistance, providing an effective solution for the field of stego-text generation. Future work will focus on utilizing fine-tuning techniques to improve generative steganography models to further enhance the security of covert communications.

\bibliographystyle{elsarticle-num} 
\bibliography{references} 

\end{document}